\documentstyle[aps,pre,multicol,epsfig]{revtex}

\begin{document}
\draft
\title{Mode Fluctuation Distribution for Spectra of Superconducting Microwave 
Billiards}
\author{H. Alt$^{1}$, A. B\"acker$^{2}$, C. Dembowski$^{1}$\footnote{Present 
        address: University of Colorado, Department of Physics, Boulder, 
        CO 80309-0390, USA}, 
        H.-D. Gr\"af$^{1}$, R. Hofferbert$^{1}$,
        H. Rehfeld$^{1}$, and A. Richter$^{1}$\\
       }
\address{$^{1}$
         Institut f\"ur Kernphysik, Technische Universit\"at Darmstadt,
         D-64289 Darmstadt, Germany\\
         $^{2}$
         Abteilung Theoretische Physik, Universit\"at Ulm, 
         D-89069 Ulm, Germany}
\date{\today}
\maketitle
\begin{abstract}
High resolution eigenvalue spectra of several two- and three-dimensional 
superconducting microwave cavities have been measured in the frequency range
below 20 GHz and analyzed using a statistical measure which is given by the 
distribution of the normalized mode fluctuations. For chaotic systems the limit
distribution is conjectured to show a universal Gaussian, whereas integrable 
systems should exhibit a non-Gaussian limit distribution. For the investigated 
Bunimovich stadium and the 3D-Sinai billiard we find that the distribution is
in good agreement with this prediction. We study members of the family of 
lima\c{c}on billiards, having mixed dynamics. It turns out that in this case 
the number of approximately 1000 eigenvalues for each billiard does not allow 
to observe significant deviations from a Gaussian, whereas an also measured 
circular billiard with regular dynamics shows the expected difference from a 
Gaussian. 
\end{abstract} 
\begin{multicols}{2}
\narrowtext
\newcommand{\R}{{\bf R}}
\newcommand{\bfq}{{\bf q}}
\newcommand{\CA}{{\cal A}}
\newcommand{\CL}{{\cal L}}
\newcommand{\CC}{{\cal C}}
\newcommand{\Mbox}[1]{{\mbox{\scriptsize #1}}}
\newcommand{\ud}{\mbox{\rm d}}

\section{Introduction}
\label{introduction}

One of the main research lines in quantum chaos is to investigate the 
statistics of energy levels of quantum systems whose classical counterpart is 
chaotic. A very popular class of systems are Euclidean billiards which are 
classically given by the free motion inside a domain $\Omega\subset \R^2$ with 
elastic reflections at the boundary $\partial \Omega$. The corresponding 
quantum billiard is described by the stationary Schr\"odinger equation 
($\hbar=2m=1$)
\begin{equation}
  \Delta \psi_n(\bfq) +k_n^2 \,\psi_n(\bfq)=0 
       \qquad \mbox{ for } \bfq \in \Omega
\end{equation}
with Dirichlet boundary conditions 
$\psi_n(\bfq)=0$ for \mbox{$\bfq\in\partial\Omega$}.

It has been conjectured that the energy level statistics of integrable systems 
can be described by a Poissonian random process \cite{BerTab77}, whereas 
classically strongly chaotic systems should obey the statistics of random 
matrix ensembles like the GOE or GUE \cite{McDKau79,CasValGua80,BohGiaSch84}.
This implies for example that the nearest neighbor level spacing distribution
is expected to show level repulsion for chaotic systems, in contrast to 
integrable systems which are expected to show level attraction. Surprisingly, 
this means that the statistics of classically chaotic systems are much more
rigid than those of integrable systems.

These conjectures have been tested successfully for several systems. However, 
there are exceptions for both integrable and chaotic systems. For example the 
so-called arithmetic systems, which are strongly chaotic, are found to have a 
level spacing distribution showing level-attraction similar to the Poissonian 
distribution 
\cite{AurSte89,BogGeoGiaSch92,BolSteSte92,Bol93,AurSchSte95,Sar95}.
Therefore a different statistics, the distribution of the normalized mode
fluctuations has been proposed \cite{Ste94,AurBolSte94} as a possible 
signature of quantum chaos. This statistics was first investigated in the 
integrable case for the eigenvalues of the Laplacian on a torus \cite{CheLeb91}
and later in \cite{bogoschm} the unnormalized fluctuations, possessing no 
limit distribution, have been studied for regular and chaotic billiards. For 
chaotic systems the limit distribution of the normalized mode fluctuations is 
conjectured \cite{Ste94,AurBolSte94} to show a universal Gaussian, whereas 
integrable systems should exhibit a non-Gaussian limit distribution. This 
conjecture was tested successfully for several regular and chaotic billiard 
systems in \cite{AurBolSte94,AurSchSte95,BaeSteSti95,AurBaeSte97}.

By using two-dimensional microwave cavities quantum billiards can be simulated 
experimentally \cite{stoeck_90,sridhar_91,graefprl_92,deus_95}. This is 
possible because of the equivalence of the stationary Schr\"odinger equation 
for quantum billiards and the corresponding Helmholtz equation for 
electromagnetic resonators in two dimensions.  In three dimensions the
electromagnetic Helmholtz equation is vectorial and cannot be reduced to an
effective scalar form. Thus, it is structurally different from the scalar
Schr\"odinger equation. Nevertheless the applicability of the statistical 
concepts developed in the theory of quantum chaos and random matrix theory is 
also given for such three-dimensional systems 
\cite{deus_95,altpre_96,altprl_97}. Therefore experiments with superconducting 
microwave resonators provide in general eigenvalue spectra of very high 
resolution for which an analysis of the distribution of the normalized mode 
fluctuations is interesting.

The paper is organized as follows. In Sec.~\ref{modefl} the mode fluctuation 
distribution is introduced. The experimental set-up and the measurement of the 
eigenfrequencies using superconducting microwave resonators are described in 
Sec.~\ref{experiment}. In Sec.~\ref{application} the analysis of the mode 
fluctuation distribution using the experimental data is carried out.

\section{Mode fluctuation distribution}
\label{modefl}

The analysis of the eigenvalue spectrum starts with the spectral staircase 
function 
\begin{equation}
  N(k) = \#\{ n \;|\; k_n \le k \}, 
\end{equation}
which counts the number of energy levels below a given energy $k$. The mean 
behavior of $N(k)$ is given by the generalized Weyl's law \cite{BalHil76}, 
which reads for two-dimensional billiards with Dirichlet boundary conditions
\begin{equation}
  \overline{N}(k) = \frac{\CA}{4\pi} k^2 - \frac{\CL}{4\pi} k + \CC \;\;,
\end{equation}
where $\CA$ is the area of the billiard, $\CL$ the length of the boundary, and 
$\CC$ takes curvature and corner contributions into account. For 
three-dimensional electrodynamical billiards we have
\begin{eqnarray}
\overline{N}&&(k) = \frac{\cal{V}}{3\pi^2}\cdot k^3\\
&&-\left(\frac{2}{3\pi^2}\int\frac{\ud\sigma}{\cal{R}}-\frac{1}{12\pi^2}
   \int \ud a
   \frac{(\pi-\omega)(\pi-5\omega)}{\omega}\right)\cdot k \nonumber\\
&&+\; const. \;\;,\nonumber
\end{eqnarray}
where $\cal{V}$ is the volume of the billiard, $\cal{R}$ the mean radius of 
the curvature over the surface $\sigma$ and $\omega$ is the dihedral angle 
along the edges $a$ \cite{lukosz,balian}.

In order to obtain a spectrum which is independent of the system specific 
constants, one considers the unfolded spectrum $\{x_n : = \overline{N}(k_n)\}$
\cite{porter,bohigas_84}. Consequently the unfolded energy spectrum has a mean 
level spacing of unity. The counting function for the unfolded spectrum will be 
denoted for simplicity again by $N(x)$. Thus the fluctuating part of the 
spectral staircase function is given by
\begin{equation}
  N_{\Mbox{fluc}}(x) := N(x)-\overline{N}(x) = N(x)-x \;\;.
\label{n_fluc}
\end{equation}
In the following we will assume that all spectra have been unfolded and that 
the systems are completely desymmetrized.

The normalized mode fluctuations are given by
\begin{equation}
  \label{W(x)-Def}
   W(x) := \frac{N_{\Mbox{fluc}}(x)}{\sqrt{D(x)}} \;\;,
\end{equation}
where $D(x)$ is the variance
\begin{equation} \label{temporal-variance}
  D(x):= \frac{\Xi(c)}{(c-1)\,x}  
         \int\limits_x^{cx} \left[N_{\Mbox{fluc}}(y)\right]^2 \; \ud y \;\;,
\end{equation}
with $c>1$, and $\Xi(c)$ is a correction necessary for integrable systems to 
obtain for $W(x)$ a variance of one, see \cite{AurBaeSte97} for details. The 
conjecture put forward in \cite{Ste94,AurBolSte94} can be formulated as 
follows (see \cite{AurBaeSte97}):
{\it
     For bound conservative and scaling systems the quantity $W(x)$,
     Eq.\ (\ref{W(x)-Def}), possesses a limit distribution for $x\to\infty$.
     This distribution is absolutely continuous with respect to the Lebesgue 
     measure on the real line, with a density $P(W)$ defined by
     \begin{eqnarray}
       \lim_{T\to\infty} \frac{1}{(c-1)\,T} \int\limits_T^{cT} 
           g(W(x)) \, \rho(x/T) \; \ud x \nonumber \\ 
          \label{Conjecture}
         = \int\limits_{-\infty}^\infty g(W) \, P(W) \; \ud W \;\;,
     \end{eqnarray}
     where $g(x)$ is a bounded continuous function, and $\rho(t)\ge 0$ is a 
     continuous density on $[1,c]$ with 
     $\frac{1}{c-1} \int_1^c \rho(t) \, \ud t = 1$.

     Furthermore, the limit distribution has zero mean and unit variance,
     \begin{equation} \label{1st-2nd-moment}
        \int\limits_{-\infty}^\infty W \, P(W) \; \ud W   = 0 \;\;, \quad
        \int\limits_{-\infty}^\infty W^2 \, P(W) \; \ud W = 1 \;\;.
     \end{equation} 
}
Now, the basic conjecture of \cite{Ste94,AurBolSte94} reads as follows:
{\it If the corresponding classical system is strongly chaotic, having only 
     isolated and unstable periodic orbits, then $P(W)$ is universally a 
     Gaussian, 
\begin{equation}
     P(W)=\frac{1}{\sqrt{2\pi}} \exp\left(-\frac{1}{2}W^2\right).
\label{gauss_dist}
\end{equation}
     In contrast, a classically integrable system leads to a non-Gaussian 
     density $P(W)$.
}

For large classes of integrable systems it has been proven that the limit 
distribution is not Gaussian, see 
\cite{AurBaeSte97,BleDysLeb93,KosMinSin93,Ble96:p} and references therein.

For chaotic systems this conjecture has been tested numerically in 
\cite{AurBolSte94,AurSchSte95,BaeSteSti95}, and experimentally in 
\cite{SirKoc96}. A review and a detailed comparison between chaotic and 
non-chaotic systems is given in \cite{AurBaeSte97}.
A Gaussian distribution for chaotic systems corresponds to maximum 
randomness for the fluctuations, whereas the mode fluctuation distribution for 
integrable systems is less random \cite{AurBolSte94}.

Berry's semiclassical analysis \cite{berry_a400} gives results for the 
asymptotic behavior of the saturation behavior of the spectral rigidity. In a 
similar way one can determine the asymptotic behavior of $D(x)$. For generic 
integrable billiards one has
\begin{equation}
  D(x) \sim const. \sqrt{x} \;\;.
\end{equation}
For generic classically chaotic systems with anti-unitary symmetry (e.g. 
time--reversal symmetry) one obtains
\begin{equation}
  D(x) \sim \frac{1}{2\pi^2} \ln x + const. \;\;.
\end{equation}

As discussed in \cite{AurBaeSte97} one can extend the conjecture to such 
chaotic systems, for which $N_{\Mbox{fluc}}(k)$ is modulated by a long range 
oscillation $N_{\Mbox{long}}(k)$. In this case one has to include the 
additional term in the unfolding process, 
$\{x_n : = \overline{N}(k_n) + N_{\Mbox{long}}(k_n) \}$. This procedure has 
been used for instance in the case of the truncated hyperbola billiard, where 
a prominent contribution to $N(k)$ is given by families of closed non-periodic 
orbits running into a boundary point where the curvature is discontinuous 
\cite{AurHesSte95}; see \cite{AurBolSte94,AurBaeSte97} for the result of 
$P(W)$ for this system.

The same is also necessary for the stadium billiard, where a family of 
marginally stable orbits, the bouncing ball orbits (bbo) gives rise to a strong
modulation \cite{graefprl_92,SieSmiCreLit93}. Taking the contribution of the 
bbo into account one observes excellent agreement of $P(W)$ with the Gaussian 
normal distribution \cite{AurBaeSte97}.

\section{Experiment}
\label{experiment}

For a precise test of the distribution of the normalized mode fluctuations an
accurate measurement of the resonances of all investigated microwave billiards 
is necessary. Since 1991, we have experimentally studied several two- and 
three-dimensional systems using superconduct\-ing microwave resonators of 
niobium. In Fig.~1 the shapes of some measured billiards with their dimensions 
are shown. Altogether five desymmetrized two-dimensional systems are 
investigated: a $\gamma=1.8$ Bunimovich stadium billiard 
\cite{buni,graefprl_92,altprl_95}, a circular billiard (not desymmetrized)
and three members of the family of lima\c{c}on billiards, which have first 
been theoretically studied as billiards in \cite{robnik_83}. Their boundary is 
defined as the quadratic conformal map of the unit disc onto the complex 
$w$-plane: $w=z+\lambda z^2$, where $\lambda\in[0,1/2]$ controls the 
chaoticity of the system. The billiards of the lima\c{c}on family are also 
called Pascalian snails, their shape has already been mentioned by A.\ D\"urer 
in 1525 \cite{duerer}. We also analyzed a desymmetrized three-dimensional 
Sinai billiard \cite{Sin70,primack,altprl_97,primack_dr}. 

The measurements were carried out in a LHe-bath cryostat. The billiards were 
excited up to a frequency of \mbox{20 GHz}, the upper limit of the used 
HP8510B vector network analyzer, using four capacitively coupling dipole 
antennas sitting in small holes on the niobium surface and penetrating up to a 
maximum of 1 mm into the cavity to avoid disturbances of the electromagnetic 
field inside the resonators. Using one antenna for the excitation and either 
another or the same one for the detection of the microwave signal, we were able
to measure the transmission or the reflection spectrum of the resonators. The 
spectra were taken in 10 kHz-steps and the measured resonances have quality 
factors of up to $Q\approx 10^7$ and signal-to-noise ratios of up to 
\mbox{$S/N\approx70$ dB} which made it easy to separate the resonances from 
each other and from the background even in the higher frequency range, where 
the level density strongly increases. Especially in the case of the circular 
billiard with mainly twofold degenerated resonances the advantage of using 
superconducting cavities is obvious. Due to mechanical imperfections the 
degenerated modes show a very weak splitting, but nevertheless one is able to 
resolve all resonances. As a consequence, all the important characteristics 
like eigenfrequencies and widths could be extracted with a very high accuracy 
\cite{altnucphys_93,altplb_96}. A detailed analysis of the original spectra 
yields a total number of 1000 - 1200 resonances for the 2D-billiards (about 
660 resonances for the circular billiard), and nearly 1900 resonances for the 
3D-billiard. A detailed comparison with numerical data confirms that the 
measured spectra are almost complete \cite{AltDemGraHesHofRehRicSte97:p}. These
eigenvalue sequences $\{k_1, k_2, \ldots, k_n\}$ (with $k=2\pi/c_0 f$ and $c_0$
is the speed of light) form the basis of the present test of the mode 
fluctuation distribution.

\begin{figure} [htb]
\centerline{\epsfxsize=8.6cm
\epsfbox{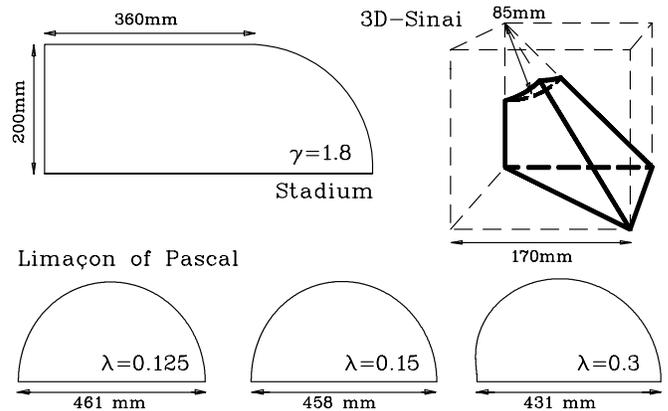}}
\vspace*{2.5ex}
\caption{Investigated billiard systems. In the upper part on the left side a 
Bunimovich stadium billiard ($\gamma=1.8$) and on the right a 3D-Sinai billiard
are illustrated. In the lower part three billiards of the lima\c{c}on family 
with different parameters $\lambda$ are shown changing the chaoticity from 
nearly regular ($\lambda=0.125$) to chaotic ($\lambda=0.3)$.}
\end{figure}

\section{Application to the measured eigenfrequency spectra}
\label{application}

In this section we study the mode fluctuation distribution discussed in 
Sec.~\ref{modefl} for the experimentally investigated systems presented in 
Sec.~\ref{experiment}. 

\subsection{2D-Bunimovich stadium billiard}

First we want to discuss the results for the $\gamma=1.8$ stadium billiard. As 
proven in \cite{buni} the stadium billiard is strongly chaotic (i.e.\ ergodic, 
mixing, K-system). Ergodicity, however, does not prevent a system from having 
a family of marginally stable periodic orbits, as long as they are of measure
zero in phase space. In the stadium billiard such a family is given by the 
bouncing ball orbits (bbo), which have perpendicular reflections at the 
straight line segments. Their contribution dominates the fluctuating part of 
the spectral staircase function \cite{graefprl_92}. Therefore the above stated 
conjecture is in its basic form not applicable (see however the refinement 
given in \cite{AurBaeSte97} and the discussion below). After unfolding the 
spectrum and calculating the distribution of the normalized mode fluctuation 
$W(x)$ according to Eq.~(\ref{W(x)-Def}) and (\ref{temporal-variance}), a 
density $P(W)$ results as shown in the left part of Fig.~2. Obviously it is not
a Gaussian, the distribution is shifted to positive $W(x)$ due to the 
existence of the bbo. In fact, in \cite{Bol97:p} it is shown using the results 
of \cite{SieSmiCreLit93} that in this case $W(x)$ is a bounded function, such 
that the limit distribution $P(W)$ is not Gaussian.

\begin{figure}
\centerline{\epsfxsize=8.6cm
\epsfbox{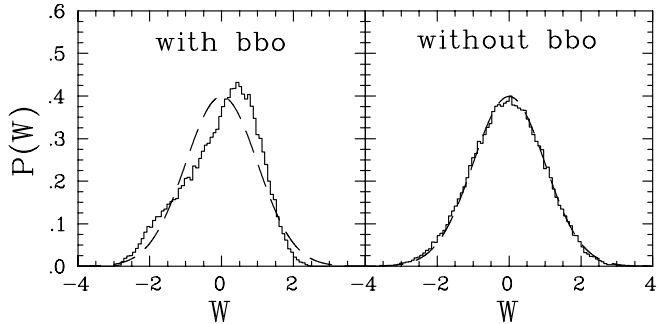}}
\vspace*{2.5ex}
\caption{Mode fluctuation distribution $P(W)$ for the $\gamma=1.8$ Bunimovich
stadium billiard. On the left side the distribution with bouncing ball orbits 
(bbo) is shown, which is asymmetric with bias to positive $W$. On the right 
side the distribution without bbo is displayed, which is in very good agreement
with the Gaussian (dashed curve) expected for chaotic systems 
(Eq.~(\ref{gauss_dist})).}
\end{figure}

Since $P(W)$ is expected to show Gaussian behavior only if the corresponding 
classical system is strongly chaotic excluding stable or neutrally stable 
motion, the contribution of the bbo has to be taken into account in the 
unfolding procedure \cite{AurBaeSte97}, see Sec.~\ref{modefl}. Finally, the 
predicted Gaussian distribution of $P(W)$ is obtained, as can be seen in the 
right part of Fig.~2. To have a measure how good the distribution $P(W)$ 
agrees with a Gaussian we use the Kolmogorov-Smirnov test, which gives a 
significance level computed from the maximum value of the absolute difference 
between the two cumulative distributions. 

For the case with bbo one obtains a significance level of $\mbox{57.2\%}$ and 
for the case with extracted bbo of $\mbox{75.2\%}$. The influence  of the bbo 
which are visible in the distributions in Fig.~2 is also reflected in the 
value of the significance level.

To determine the normalized mode fluctuations, Eq.~(\ref{W(x)-Def}), we have 
calculated the fluctuating part $N_{\Mbox{fluc}}(x)$ of the unfolded 
experimental spectra according to Eq.~(\ref{n_fluc}). For the variance $D(x)$, 
Eq.~(\ref{temporal-variance}), we have used the ansatz 
\mbox{$D(x)=\sqrt{const. \sqrt{x}}$} for bbo-contribution included  and 
\mbox{$D(x)=1/(2\pi^2)\ln (x)+ const.$} after their extraction, see 
Sec.~\ref{modefl}. Then $W(x)$ was calculated for $10^5$ to $10^6$ randomly 
distributed values of $x$ in the interval $[0,x_{\Mbox{max}}]$, with 
$x_{\Mbox{max}}$ being the upper unfolded eigenenergy. The constant term in 
$D(x)$ was chosen according to Eq.~(\ref{1st-2nd-moment}) to give a unit 
variance for the obtained distribution $P(W)$.

These results using the experimentally determined set of energy levels confirm 
our previous analysis using the numerically computed eigenspectrum 
\cite{AurBaeSte97}.

\subsection{3D-Sinai billiard}

Another investigated system where the bouncing ball orbits play an important 
role is the 3D-Sinai billiard, which is also proven to be strongly chaotic 
\cite{Sin70}. In contrast to the 2D-Bunimovich stadium billiard the 3D-Sinai 
billiard possesses families of bbo of dimension two and three. In Fig.~3 the 
mode fluctuation distributions with and without bbo are displayed. In the 
distribution with bbo (left part of Fig.~3) only slight deviations from the 
Gaussian occur, in particular a significant peak at $W\approx -3$ appears. Thus
the influence of the bbo is less visible in the considered energy range as in 
the case of the distribution for the $\gamma=1.8$ stadium billiard which might 
be due to the superpositions of different bbo. This confirms our previous 
results of \cite{altprl_97}.Taking into account the contribution of the bbo 
modes one obtains the right part of Fig.~3. Calculating the significance level 
for the first case (with bbo) one obtains a value of $\mbox{78.5\%}$ and for 
the second case (with extracted bbo) of $\mbox{75.2\%}$.

\begin{figure}
\centerline{\epsfxsize=8.6cm
\epsfbox{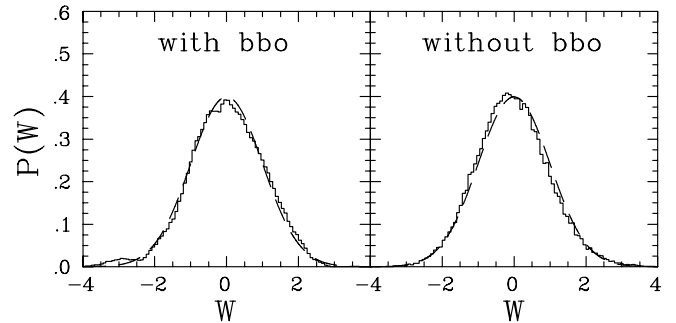}}
\vspace*{2.5ex}
\caption{Mode fluctuation distribution $P(W)$ for the 3D-Sinai billiard. On 
the left side the distribution with bouncing ball orbits (bbo) is shown, where 
at $W\approx -3$ a significant deviation from the Gaussian occurs. On the right
side the distribution without bbo is displayed, which shows the predicted 
Gaussian behavior. Again the dashed curve is the expected Gaussian for chaotic 
systems (Eq.~(\ref{gauss_dist})).}
\end{figure}

\subsection{Mixed 2D-lima\c{c}on billiards}
In this section we test the conjecture stated in Sec.~\ref{modefl} for the 
billiards of the lima\c{c}on family. We have investigated three billiards of 
different chaoticity with para\-meters $\lambda=0.125,\ \lambda=0.15 
{\rm\ and\ } \lambda=0.3$. Investigations of the classical Poincar\'e surface 
of section for these configurations have shown 
\cite{hrdipl,RehAltDemGraHofRic97:p,richterima_97} that the fraction of the 
chaotic phase space is $\mbox{55\%}$ ($\lambda=0.125$), $\mbox{66\%}$ 
($\lambda=0.15$) and nearly $\mbox{100\%}$ ($\lambda=0.3$). The quantum 
mechanical counterpart of these three billiards exhibit the same behavior 
concerning their chaoticity \cite{BerRob84,ProRob93a}, see also 
\cite{richterima_97}.

Therefore they are very suitable to study the conjecture. Note that the 
classical dynamics of the $\lambda=0.3$ billiard is not completely ergodic, 
since small stability islands still exist \cite{HDABunpublished}, see also 
\cite{hayli} for analytical results around $\lambda=0.25$. However the 
corresponding islands of stability are so small that they do not affect the 
energy spectrum in the range under consideration.

\begin{figure}
\centerline{\epsfxsize=8.6cm
\epsfbox{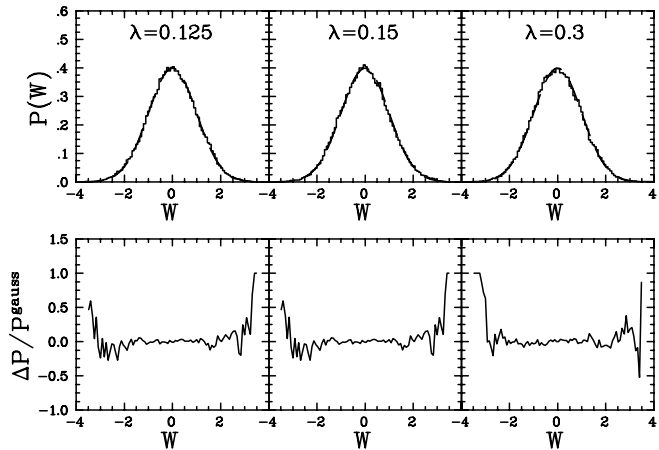}}
\vspace*{2.5ex}
\caption{Mode fluctuations distributions $P(W)$ for the three investigated
billiards of the lima\c{c}on family in the upper part. All three distributions 
follow the predicted Gaussian for chaotic systems which is represented by the 
dashed curve (Eq.~(\ref{gauss_dist})). In the lower part the differences 
between the histogram and a Gaussian is plotted.}
\end{figure}

As can be seen in Fig.~4 the billiard with $\lambda=0.3$ shows indeed the 
predicted Gaussian distribution $P(W)$. For the calculation of $P(W)$ we 
followed the same procedure described for the stadium billiard with an ansatz 
for the variance \mbox{$D(x)=1/(2\pi^2)\ln (x)+ const.$}\ . On the other hand 
the two billiards with $\lambda=0.125$ and $\lambda=0.15$ belong to the class 
of mixed systems. Their classical counterparts possess aside from isolated and 
unstable periodic orbits also stable orbits \cite{robnik_83,richterima_97}. 
From this one would expect, that according to this the distribution $P(W)$ 
should show non-Gaussian behavior. However, the histograms in Fig.~4, obtained 
with assuming $D(x)\propto \sqrt{x}$, which gives a good description for 
$D(x)$ in the considered enery interval, allow no significant distinction 
between $P(W)$ and the Gaussian distribution, see also the lower part of 
Fig.~4, where the difference $\Delta P$ between a Gaussian and the calculated 
distribution $P(W)$ is shown. These characteristics are also expressed in the 
significance levels obtained from the Kolmogorov-Smirnov test which lie around 
$\mbox{85\%}$. Presumably one needs a large number of energy levels to be able 
to see significant deviations.

\subsection{2D-Circular billiard}

Finally we have studied the mode fluctuation distribution for the circular 
billiard. Since this system is integrable, one would expect a deviation of the 
mode fluctuation distribution from a Gaussian due to the existence of neutrally
stable periodic orbits. To obtain the distribution $P(W)$ we use the ansatz 
for the variance $D(x)=const. \cdot \sqrt{x}$. In Fig.~5 the distribution 
$P(W)$ is shown, which  clearly deviates from the Gaussian. For the circular 
billiard one obtains a significance level of $67.1\%$.

This result is in contrast to the result of \cite{SirKoc96}, where no 
difference between the mode fluctuation distribution for the numerical obtained
eigenmodes of a regular system and a Gaussian for chaotic systems could be 
found, although nearly the same number of eigenfrequencies ($\approx 660$) are 
used.

\begin{figure}
\centerline{\epsfxsize=8.6cm
\epsfbox{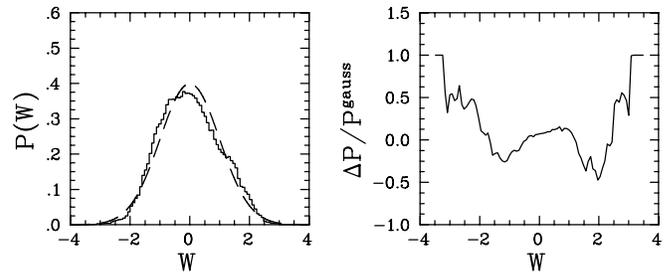}}
\vspace*{2.5ex}
\caption{Mode fluctuation distribution $P(W)$ for the circular billiard on the 
left side. The histogram shows the expected deviation from a Gaussian for 
chaotic systems which is represented by the dashed curve 
(Eq.~(\ref{gauss_dist})). On the right side one can see the difference between 
the Gaussian and the histogram.}
\end{figure}
           
\section{Conclusion}
\label{conclusion}

In this paper we have studied the distribution of the normalized mode 
fluctuations for experimentally obtained  high resolution eigenvalue spectra of 
several two- and three-dimensional superconducting microwave billiards. The 
conjecture for this statistical measure states that the distribution $P(W)$ of 
the normalized mode fluctuations for a given eigenvalue spectrum leads to a 
Gaussian when the corresponding classical system is strongly chaotic having 
only unstable and isolated periodic orbits.

This has been successfully tested using the eigenvalues of a $\gamma=1.8$
stadium billiard and a 3D-Sinai billiard. Both systems are strongly chaotic, 
but possess families of bouncing ball orbits, whose contribution has been 
subtracted for the determination of $P(W)$. The same result has been obtained 
with a $\lambda=0.3$ billiard of the lima\c{c}on family. Two other investigated
billiards of the lima\c{c}on family, the $\lambda=0.125$ and $\lambda=0.15$ 
billiard, belong to the class of mixed systems. The fact that no visible 
difference from a Gaussian occurs in the distribution $P(W)$ might be due to 
the finite number of eigenvalues in the given energy range we used to 
calculate the distribution and the chaoticities of the systems (55\% resp. 
66\%) which are closer to chaotic than to regular dynamics. A variation of the 
included number of modes (500, 600, ..., 1100) in the mode fluctuation 
distribution shows no significant change in the results. In the case of the 
circular billiard we get a clear deviation from the Gaussian, as one would 
expect. This is in contrast to the result of \cite{SirKoc96}, where no such 
difference for a regular system could be found.

Therefore characterizing the chaoticity using a small number of energy levels 
with the help of the conjecture stated in \cite{AurBolSte94} is very difficult.
For regular resp. chaotic systems the reachable number of eigenvalues obtained 
by experiments explained in Sec.~\ref{experiment} is sufficient to have 
satisfactory results for the mode fluctuation distribution. But for special 
regular (rectangular billiards \cite{SirKoc96}) or mixed systems the needed 
number of eigenvalues, approximately $10^4 - 10^6$, can experimentally only 
hardly be achieved, so that a practical usage of this conjecture to obtain 
informations about the chaoticity of such systems is limited. 

\section{Acknowledgments}
We would like to thank H. Lengeler and the mechanical workshop at CERN/Geneva
for the precise fabrication of the niobium resonators. A.B.\ is grateful to 
R.\ Aurich, T.\ Hesse, R.\ Schubert and F.\ Steiner for useful discussions and 
comments, and one of us (A.R.) acknowledges advice given by E.\ Bogomolny.
This work has been supported by the Sonderforschungsbereich 185
``Nichtlineare Dynamik'' of the Deutsche Forschungsgemeinschaft and in part by 
the Bundesministerium f\"ur Bildung und Forschung under contract number 06DA820.
A.B.\ acknowledges support by the Deutsche Forschungsgemeinschaft under 
contract No.\ DFG-Ste 241/7-2.

\end{multicols}
\end{document}